\documentclass[%
 aip,
 amsmath,amssymb,
]{revtex4-1}

\usepackage{graphicx}
\usepackage{dcolumn}
\usepackage{bm}

\usepackage[utf8]{inputenc}
\usepackage[T1]{fontenc}
\usepackage{mathptmx}
\usepackage{etoolbox}

\makeatletter
\def\@email#1#2{%
 \endgroup
 \patchcmd{\titleblock@produce}
  {\frontmatter@RRAPformat}
  {\frontmatter@RRAPformat{\produce@RRAP{*#1\href{mailto:#2}{#2}}}\frontmatter@RRAPformat}
  {}{}
}%
\makeatother
\begin{document}

\preprint{Arxiv version}

\title[]{Shock induced  aerobreakup of a polymeric droplet}
\author{Navin Kumar Chandra}
\author{Shubham Sharma}%
 \affiliation{Department of Mechanical Engineering, Indian Institute of Science, Bangalore- KA560012, India}

\author{Saptarshi Basu*}
 
 \email{sbasu@iisc.ac.in}
\affiliation{Department of Mechanical Engineering, Indian Institute of Science, Bangalore- KA560012, India}
\affiliation{Interdisciplinary Centre for Energy Research, Indian Institute of Science, Bangalore- KA560012, India}

\author{Aloke Kumar*}
 
 \email{alokekumar@iisc.ac.in}
\affiliation{Department of Mechanical Engineering, Indian Institute of Science, Bangalore- KA560012, India}

\begin{abstract}
Droplet atomization through aerobreakup is omnipresent in various natural and industrial processes. Atomization of Newtonian droplets is a well-studied area; however, non-Newtonian droplets have received less attention despite their frequent encounters. By subjecting polymeric droplets of different concentration to the induced airflow behind a moving shock wave, we explore the role of elasticity in modulating the aerobreakup of viscoelastic droplets. Three distinct modes of aerobreakup are identified for a wide range of Weber number $(\sim10^2-10^4)$ and Elasticity number $(\sim10^{-4}-10^2)$ variation; these modes are- vibrational, shear-induced entrainment and catastrophic breakup mode. Each mode is described as a three stage process. Stage-I is the droplet deformation, stage-II is the appearance and growth of hydrodynamic instabilities, and stage-III is the evolution of liquid mass morphology. It is observed that elasticity plays an insignificant role in the first two stages, but a dominant role in the final stage. The results are described with the support of adequate mathematical analysis. 
\end{abstract}

\maketitle

\section{Introduction}
\label{sec:Introduction}

Secondary atomization is the process of breaking a liquid droplet into smaller units. Aerobreakup is one example of secondary atomization in which a liquid droplet is exposed to a high-speed stream of gas (generally air), causing its fragmentation. Aerobreakup applies in various natural and industrial processes. Mixing of air and fuel droplets inside an internal combustion engine, gelled propellants in rocket engine \citep{padwal2021gel}, breakup of sneezed salivary droplets \citep{scharfman2016visualization}, falling raindrops \citep{villermaux2009single}, and powder production by spray atomization of fruit pulps \citep{cervantes2014study} etc. are few instances that involve aerobreakup of liquid droplets. Understanding the physics of aerobreakup is crucial in designing and controlling these processes. Significant research has been already done to study the aerobreakup of Newtonian droplets, and they are well-reviewed in literature \citep{pilch1987use,gelfand1996droplet,guildenbecher2009secondary,jackiw2021aerodynamic,sharma2021shock}. It is well established that the two most important dimensionless groups in the study of Newtonian droplet aerobreakup are Weber number ($We$) and Ohnesorge number ($Oh$), defined as
\begin{equation}
We=\frac{\rho_g U_g^2 D_0}{\gamma} \hspace{0.2cm}; \hspace{1cm} 
Oh=\frac{\mu_l}{\sqrt{\rho_l \gamma D_0}}
\label{eqn:eqn0}
\end{equation}
$\rho_g$ and $\rho_l$ are the gas and liquid phase density, $U_g$ is the free stream velocity of the gas, $\gamma$ is the surface tension at the liquid-gas interface, $D_0$ is the initial diameter of the droplet and $\mu_l$ is the dynamic viscosity of the liquid phase. Historically, based on the morphology, droplet breakup has been categorized into five modes generally represented on a $We-Oh$ plane and occurs at increasing order of Weber number. These modes are: Vibrational, bag, multi-mode (including bag and stamen), shear stripping or sheet thinning and catastrophic mode of breakup \citep{guildenbecher2009secondary,jain2015secondary}.
Aerobreakup of a non-Newtonian droplet differs significantly from Newtonian droplets \citep{matta1982viscoelastic,arcoumanis1994breakup,wilcox1961retardation,theofanous2013physics}. Despite their occurrence in many practical processes, research in the aerobreakup of non-Newtonian droplets is still lacking. It has been pointed out that consensus on something as basic as the breakup modes and even the suitable dimensionless groups are not clear for the non-Newtonian liquids \citep{guildenbecher2009secondary,guildenbecher2011droplet}. In several studies, polymeric solutions have been employed as the model fluid to investigate the aerobreakup of non-Newtonian liquids \citep{wilcox1961retardation,hoyt1980drag,matta1982viscoelastic,matta1983aerodynamic,joseph2002rayleigh,theofanous2013physics}. Long-chain polymer molecules impart elasticity when dissolved into viscous solvents, and the resulting solution exhibit viscoelastic behavior.\\
Early research in the area of the polymeric droplet breakup was focused primarily on the resultant fragment size of liquid mass \citep{wilcox1961retardation,matta1982viscoelastic,matta1983aerodynamic}. These studies outlined the role of elasticity as the retardation to the breakup process in two aspects- larger fragment size and higher breakup time when compared with the results of Newtonian (viscous) droplet under similar conditions. Later on, the focus of the research was shifted towards identifying the breakup modes and underlying mechanism for aerobreakup of viscoelastic droplets \citep{arcoumanis1994breakup,joseph1999breakup,joseph2002rayleigh,ng2008modes,theofanous2011aerobreakup,theofanous2013physics,theofanousMitkin2017physics}. \citet{arcoumanis1994breakup} noted that the aerobreakup always starts with the appearance of waves on the droplet surface. In case of the polymeric droplet, these waves evolve into long ligaments and finally detach from the primary droplet but not in the form of daughter droplet like Newtonian fluids. At very high Weber numbers ($\sim 10^4$), droplets undergo a widespread catastrophic breakup marked at early times by the appearance of surface corrugations in the droplet frontal area. A match between experiments and theory for both, Newtonian and viscoelastic liquids suggested that the surface corrugations are  Rayleigh-Taylor (RT) waves \citep{joseph1999breakup,joseph2002rayleigh}. However, there is ambiguity in the existence of the RT waves and the catastrophic breakup mode itself at high Weber number \citep{theofanous2008physics,theofanous2011aerobreakup,theofanous2013physics}. Advancements in experimental facilities like high-speed cameras with better resolution, pulsed laser with nanosecond accuracy and Laser-induced fluorescence imaging technique etc. re-ignited the research in the field of aerobreakup \citep{sharma2021shock,theofanous2008physics}. The possibility of unifying Newtonian and non-Newtonian breakup modes under a single roof has been explored \citep{theofanous2011aerobreakup}. In this endeavor, the RT and the Kelvin-Helmholtz (KH) instabilities were identified as the critical mechanism that decides the mode of aerobreakup. The respective modes due to these instabilities are Rayleigh-Taylor piercing (RTP) and shear-induced entrainment (SIE). A third mode, shear-induced entrainment with rupture (SIER) has also been proposed only for viscoelastic liquids \citep{theofanous2013physics}. It is not always the case that only one of the two instabilities (RT and KH) will govern the breakup process, modulation of the two is also possible as reported in the secondary atomization of coal water slurry and shear-thickening viscoelastic droplets \citep{theofanousMitkin2017physics,zhao2014influence,sharma2021shock}. 

Most of the existing literature agree with the overall role of liquid elasticity as a retarding agent to the aerobreakup process. However, clear elucidation of the exact mechanism by which elasticity enters into the play is missing. Some noteworthy efforts has been made by \citet{joseph2002rayleigh} and \citet{theofanous2013physics}, but not complete in all aspects. \citet{theofanous2013physics} considered RT and KH instabilities as the main physics that decides the breakup mode; but role of elasticity in modulating these governing instabilities needs investigation. \citet{joseph2002rayleigh} studied the role of elasticity only on RT instability observed at high Weber numbers, but the proposed theory is in disagreement with the widely reported experimental observation of elasticity as a retarding agent to the aerobreakup.

Here we explore the role of elasticity and the mechanism by which, it modulates the aerobreakup of viscoelastic droplets. Aqueous solutions of Polyethylene oxide (PEO) have been employed as the model viscoelastic fluid. Variation in elastic properties is achieved by changing the polymer concentration, and quantified in terms of the Elasticity number, $El$ which is the ratio of Deborah number, $De$ and the Reynolds number, $Re_l$ in the liquid phase, such that
\begin{equation}
El= \frac{De}{Re_l}  =\frac{\lambda\mu_0}{\rho_lD_0^2}
\label{eqn:eqn00}
\end{equation}
Where, $\lambda$ and $\mu_0$ are the relaxation time and the zero-shear viscosity of the polymeric solution. A wide range of $El$ $(\sim10^{-4}-10^2)$ and $We$ $(\sim10^2-10^4)$ is investigated while keeping the droplet diameter ($\sim1.8$ mm) fixed. Three distinct breakup modes- vibrational, shear-induced entrainment (SIE) and catastrophic mode, are identified with increasing $We$ on a $We-El$ number plane. Based on the temporal evolution of liquid mass, we describe each breakup mode as a three stage process. Dominant role of liquid elasticity is observed only in the final stage of breakup, whereas it plays an insignificant role in the first two stages. Present study outlines the role of the liquid elasticity on the underlying mechanism for each stage.

\section{Materials and methods}
\label{sec:Materials and methods}
The aerodynamic breakup of a polymeric droplet is achieved in the present work by interacting an acoustically levitating droplet with a uniform stream of induced airflow generated behind a normal shock wave. The studied mechanism involves two stages \citep{sharma2021shock}. The first stage corresponds to the interaction of a shock wave with the droplet where different shock structures (such as reflected wave, transmitted wave, Mach stem, slip surface, etc.) are formed. However, in our earlier work \citep{sharma2021shock} it has been shown that this stage has negligible influence on droplet deformation and breakup dynamics. Therefore shock wave dynamics is not discussed in this work. The second stage involves shock-induced airflow interaction with the droplet, which influences the droplet's deformation and breakup. Present work is focused only on the second stage.

\begin{figure*}
\centering
\includegraphics[width=1\linewidth]{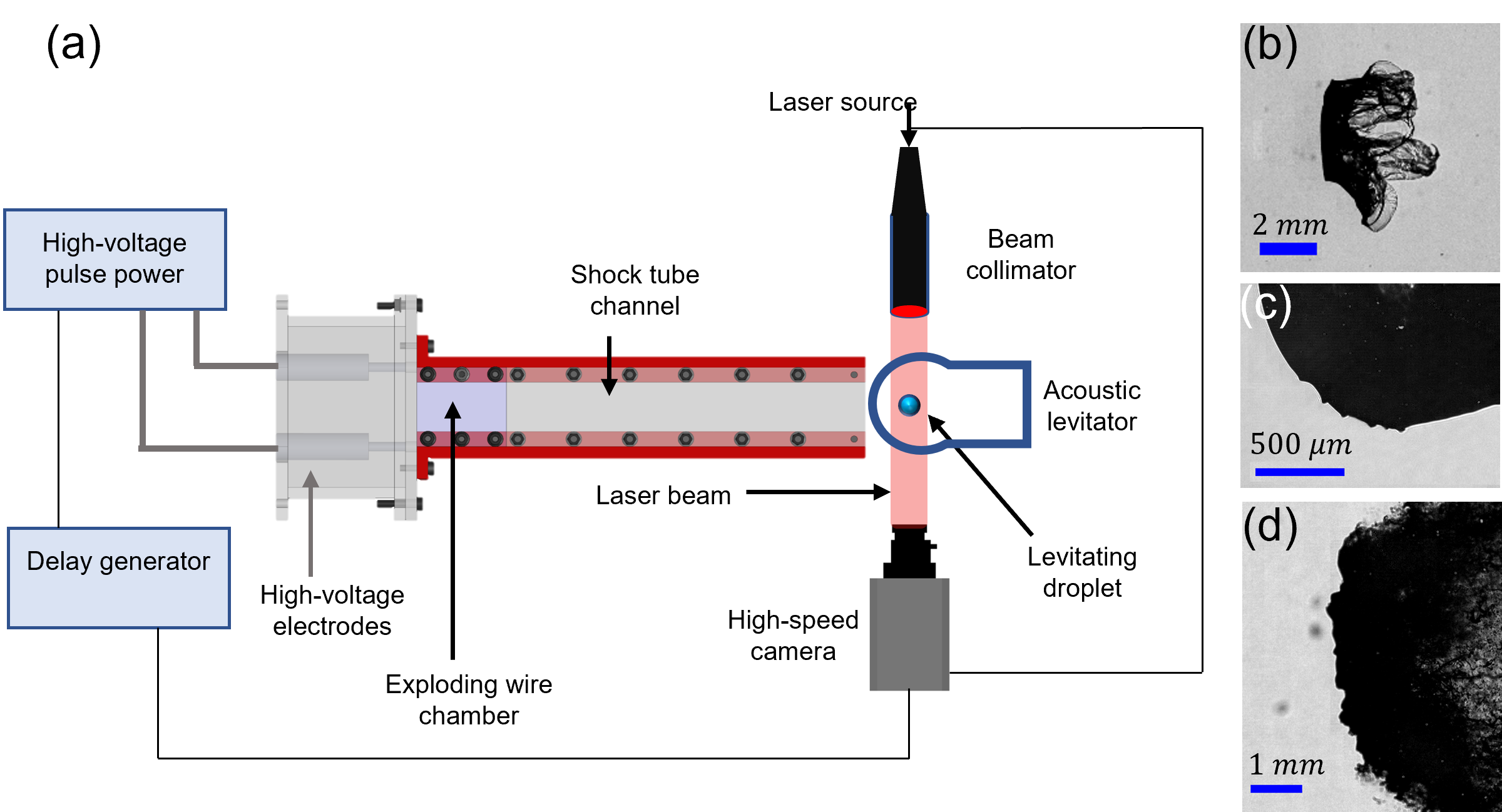}
\caption{(a) Schematic of the experimental setup. (b)-(d) Sample images showing the three different zoom settings used in the present study. (b) Zoomed out imaging for global observation. (c) Imaging with very high magnification on quarter portion of a droplet to capture the KH waves. (d) Medium zoom to capture the RT waves on the flattened frontal area of the droplet. }
\label{fig:expsetup}
\end{figure*}

\subsection*{Shock tube setup}
\label{subsec:Shock tube setup}
\begin{figure*}
\centering
\includegraphics[width=1\linewidth]{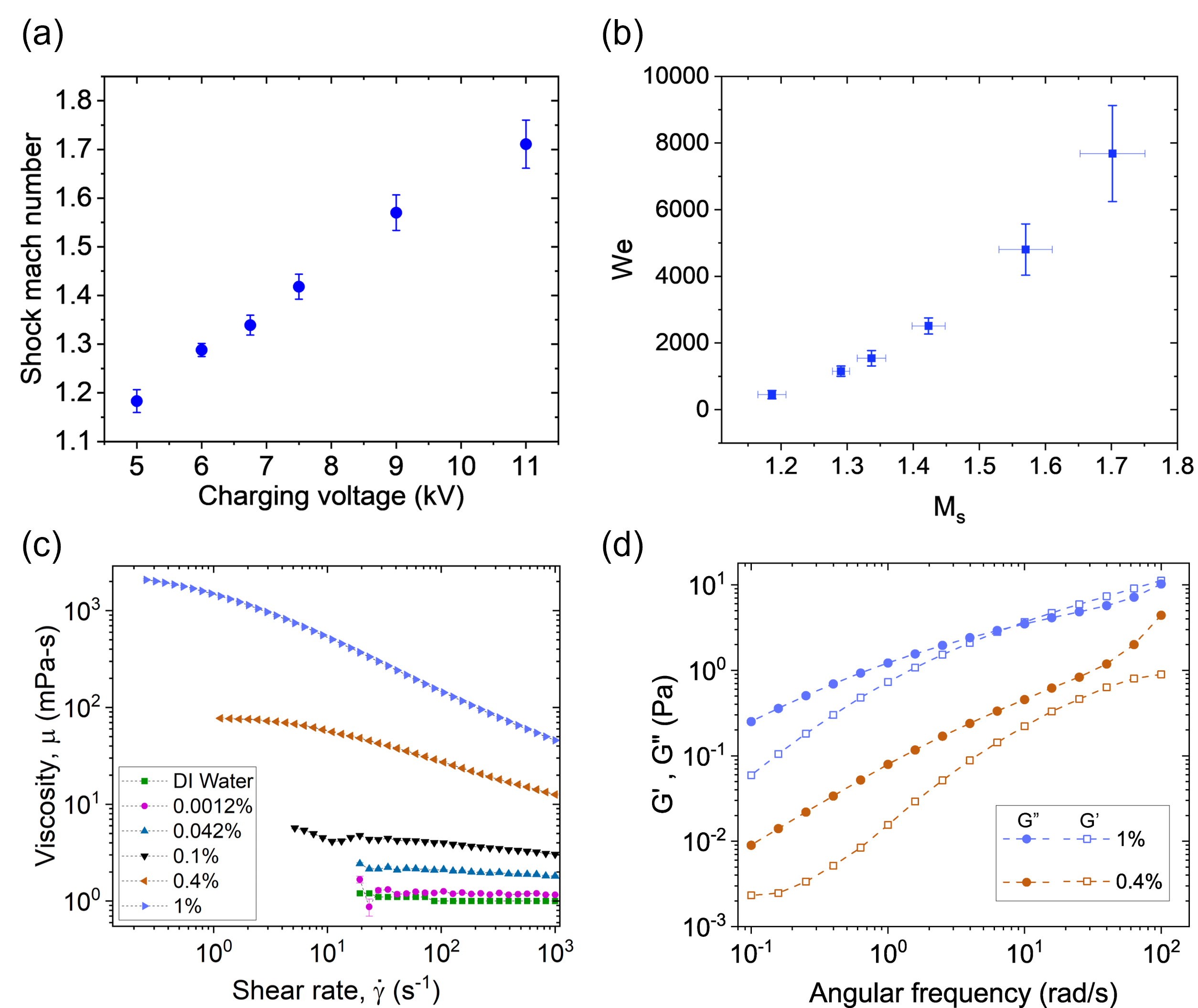}
\caption{Range of non-dimensional numbers and rheological properties of PEO-water solution at different concentrations. (a) Shock Mach number ($M_s$) vs capacitor charging voltage of pulse power system. (b) Weber number($We$) vs Shock Mach number($M_s$) range for the studied conditions. (c) Viscosity variation with shear rate for different concentration of PEO-water solution. (d) Variation of storage modulus $(G')$ and loss modulus $(G'')$ with angular frequency in SAOS test for two different concentrations of PEO-water solution. }
\label{fig:characterization}
\end{figure*}

An exploding-wire-based shock tube setup creates a shock-induced airflow that interacts with a levitating droplet. The operation of the shock tube is based on the passage of high-Ampere( in order of kilo-amperes) and high voltage (in order of kilo-volts) electrical pulse through a thin metallic wire (35 SWG, bare copper wire) mounted on two high-voltage electrodes (see figure \ref{fig:expsetup}a). The deposition of high electrical power in short time duration (order of microseconds) results in the rapid Joule heating of the wire, its instant melting, and vaporization into a column of dense vapors \citep{sembian2016plane,sharma2021shock}. The expansion of this vapor column results in forming a cylindrical shock wave. This cylindrical shock wave gets transformed into a normal shock by the rectangular confinement of the shock tube flow channel (360 mm $\times$ 50 mm $\times$ 20 mm). A 2 kJ pulse power system (Zeonics Systech, India Z/46/12) that discharges a 5 $\mu$F capacitor is used to provide a high-voltage pulse across the exploding wire. The charging voltage for the capacitor is varied from 5 $kV$ to 11 $kV$, causing the generation of shock waves with different strengths. The shock Mach number $(M_s=U_s/v)$ of produced shock waves ranges from $\sim$1.13 to 1.76 which results in a wide range of Weber number variation ($\sim10^2 - 10^4$) as presented in Figure \ref{fig:characterization}a and \ref{fig:characterization}b. Here, $U_s$ is the shock speed at interaction instant with the droplet and is measured using the distance moved by the shock wave in two consecutive camera frames. $v$ is the speed of sound in the medium ahead of the shock wave, i.e., air at 1 atm and 298 K in the present case. A detailed overview of exploding wire technique and its application in shock-wave generation can be found in these references: \citet{sharma2021shock,sembian2016plane,liverts2015mitigation,fedotov2010extreme}. In comparison with the diaphragm-based conventional shock tubes, the present technique provides several advantages such as a small size test facility, ease of operation, extensive range of shock Mach numbers ($M_s =$ 1 to 6 )\citep{sembian2016plane} and high repeatability between the tests.

BIGlev acoustic levitator \citep{marzo2017tinylev} is used in the present work to levitate the polymeric/water droplets of size $D_o \sim 1.8$ mm. The levitator is composed of ultrasonic transducers of size 16 mm and an operating frequency of 40 kHz. An array of 36  transducers is mounted on each of two curved plates separated by a designed distance for creating standing acoustic waves. The standing wave will have stable nodes at which a liquid droplet can be firmly trapped. The sound pressure level (SPL) acting on the droplet is varied by changing supplied DC voltage. High SPL is used during droplet deployment, which is then reduced until the droplet takes a spherical shape. Although care has been taken to maintain the spherical shape of a levitated droplet, some flattening of the droplet surface occurs due to the acoustic pressure acting on the droplet surface. The maximum aspect ratio ($A.R. = D_{max}/D_{min}$ ) of the levitated droplet prior to shock interaction is found to be 1.2. Here, $D_{max}$ and $D_{min}$ are the maximum and minimum diameters of a fitted ellipse on the droplet periphery and are obtained using the "Analyse particle" plugin in the $ImageJ$ software. The equivalent droplet diameter is therefore obtained as $D_o = \sqrt[3]{{D_{max}}^2\times D_{min}}$ .

The connection diagram of different types of equipment used for the operation of shock tube setup is shown in figure \ref{fig:expsetup}a. A test fluid droplet is first trapped in the stable node of the acoustic levitator. Two high voltage electrodes are connected with the pulse power system using high-tension wires. For achieving a wire explosion, a 5 $\mu$F capacitor in the pulse power system is charged to desired energy level depending on the required shock strength. Once the charging of the capacitor is complete, the charging circuit is cut off, and the capacitor's discharging circuit is closed by providing a trigger signal from a digital delay generator (BNC 575) unit. This results in the generation of a shock wave that travels along the shock tube's length and interacts with the levitating droplet. The droplet is centrally positioned to the shock tube cross-section at a distance of 15 mm from the exit. A simultaneous trigger signal is also provided to the imaging setup for the synchronized recording of the interaction phenomenon.

\subsection*{Imaging setup}
\label{sec:Imaging setup}

The droplet aerobreakup process involves a multitude of length and time scales. A high-speed camera (Photron SA5) synchronized with an ultra-high-speed pulsed nanosecond laser (Cavitar Cavilux smart UHS) allowed us to freeze the interaction phenomenon in the 10-40 ns time scale. The motion freezing avoids the streaking of high-speed droplet fragments, which otherwise might lead to observational errors. The breakup process is captured using shadowgraphy imaging technique (see Figure \ref{fig:expsetup} a). The diverging light from a high-speed laser is fed to a beam collimator (Thorlabs, BE20M-A), transforming it into a parallel light beam, resulting in the uniform illumination of the camera field of view. Acoustically trapped droplet is kept in the path of the parallel light beam, which results in the projection of droplet shadow on the camera sensor.

Side-view images are obtained at different zoom settings to capture the different aspects of droplet breakup, as shown in Figure \ref{fig:expsetup}b,c, and d. Zoomed-out imaging is performed to make a global observation on the evolution of droplet morphology by keeping the droplet in the frame for a longer duration, as shown in Figure \ref{fig:expsetup}b. A macro lens (Sigma DG 105 mm) coupled with a high-speed camera is used for zoom-out imaging. The interaction dynamics is captured at 40000 frames per second (fps) with a frame size of $640 \times 264$ pixels. A pixel resolution of 43.1 $\mu$m/pixel is obtained which result in the field of view of 27.6 mm $\times$ 11.4 mm. 

Zoomed-in imaging is done on the quarter portion of a droplet at a very high spatial resolution to capture the evolution of KH waves appearing on the droplet surface (see Figure \ref{fig:expsetup}c). The micron-size surface corrugations appear within a few microseconds of shock wave interaction, which necessitate the usage of an imaging system with high spatio-temporal resolution. A Navitar 6.5x zoom lens with a 1.5X objective and 1x adapter tube is coupled with the high-speed camera, which captures the growth of KH waves at an imaging rate of 75000 fps. A frame size of 320 $\times$ 264 pixels is used with a spatial resolution of 4.5$\mu$m/pixel which result in a field of view of  1.4 mm $\times$ 1.2 mm.

A medium zoomed imaging is performed to capture the RT waves, which appear as surface corrugations only after the windward side of the droplet has been sufficiently flattened by the aerodynamic forces, as shown in Figure \ref{fig:expsetup}d. A Navitar 6.5x zoom lens with a 1x adapter tube is used for capturing medium zoom images at 25000 fps. A pixel resolution of 15.9 $\mu$m/pixel and frame size of  640 $\times$ 448 pixels is used, resulting in a field of view of  10.2 mm $\times$ 7.1 mm.  

Airflow direction is from left to right in all the experimental images presented in this article. The usage of high-quality and precise experimental arrangements and the wide range of non-dimensional numbers covered in the present work provides an excellent benchmark for future numerical and experimental studies.

\subsection*{Sample preparation and characterization }
A known quantity of polyethylene oxide (PEO) with viscosity averaged molecular weight $5\times10^6$ g/mol purchased from Sigma-Aldrich, is dissolved into DI water to prepare the polymeric solution of desired concentration, $c$. The solutions are prepared by using a magnetic stirrer set to rotate at 300 RPM. The water-polymer mixture is stirred till an optically transparent solution is obtained. The duration of stirring depends upon the concentration of the solution and the maximum duration is $\sim$72 hours for the maximum concentration (1$\%$ w/w) considered in the present study. The value of critical overlap concentration, $c^*\approx0.06\% w/w$ and relaxation times are estimated from the correlations available in the literature \citep{varma2021coalescence}. A wide range of $c/c^*$ variation ($\sim10^{-2}-10^1$) is considered to cover from the dilute to the concentrated entangled regime of polymeric solution. This resulted in several orders of magnitude variation in $El$ ($\sim10^{-4}-10^2$). It should be noted that, since there is no variation in droplet size, $c/c^*$ and $El$ both represents the degree of elasticity and they have been used interchangeably throughout this article.

Surface tension of DI water and polymeric solutions are measured using an optical contact angle measuring and contour analysis system (OCA25) instrument from DataphysicsVR by the pendant drop method. Rheological measurements are performed using cone and plate geometry (plate diameter: 40 mm; cone angle: 1$\deg$) of a commercial rheometer (Anton Paar, model: MCR302). A concentric cylinder geometry (cylinder diameter: 39 mm) having higher sensitivity compared to cone and plate geometry is used for rheological characterization of water and dilute solutions. Flow curve for different solutions is shown in Figure \ref{fig:characterization}c, and Figure \ref{fig:characterization}d shows the storage modulus $(G')$ and the loss modulus $(G'')$ obtained from frequency sweep of small amplitude oscillatory shear (SAOS) test at a strain amplitude of 10$\%$. Solutions with concentration 1$\%$ and 0.4$\%$ w/w falls in the entangled regime of the polymeric solution \citep{varma2022rheocoalescence} and shows significantly high zero-shear viscosity compared to the remaining solutions. These higher concentration solutions also exhibit significant amount of elasticity which can be inferred from their comparable magnitude of $G'$ and $G''$. Determination of $G'$ and $G''$ from SAOS for lower concentrations is beyond the resolution of the present rheometer therefore the data is shown only for higher concentrations in Figure \ref{fig:characterization}d. It is interesting that, although the polymeric solutions with low concentrations (0.0012$\%$ and 0.042$\%$ w/w) exhibit properties similar to that of DI water in shear rheology, and yet they show significant differences in aerobreakup (as we present in the later sections). A summary of different fluid properties are presented in Table\ref{Table_1}.
\begin{table}
  \begin{center}
\def~{\hphantom{0}}
  \begin{tabular}{lcccc}
      Concentration  & $c/c^*$  &   $\mu_0$ & $\lambda$ & $\gamma$ \\
        (\% w/w) &  & (mPa-s) & (ms) & (mN/m)\\[6pt]
       0 (water) & 0 & 1 & 0 & 72\\
       0.0012 & 0.02 & 1.7 & 1.5 & 62\\
       0.042 & 0.70 & 2.5 & 1.5 & 62\\
       0.1 & 1.67 & 5.7 & 1.9 & 62\\
       0.4 & 6.67 & 77 & 76 & 62\\
       1 & 16.67 & 2081 & 507 & 62 \\
  \end{tabular}
  \caption{Properties of the test liquids.}
  \label{Table_1}
  \end{center}
\end{table}

\section{Results and discussions}
\label{sec:Results and discussions}
\subsection*{Modes of droplet aerobreakup}
\begin{figure*}
\centering
\includegraphics[width=1\linewidth]{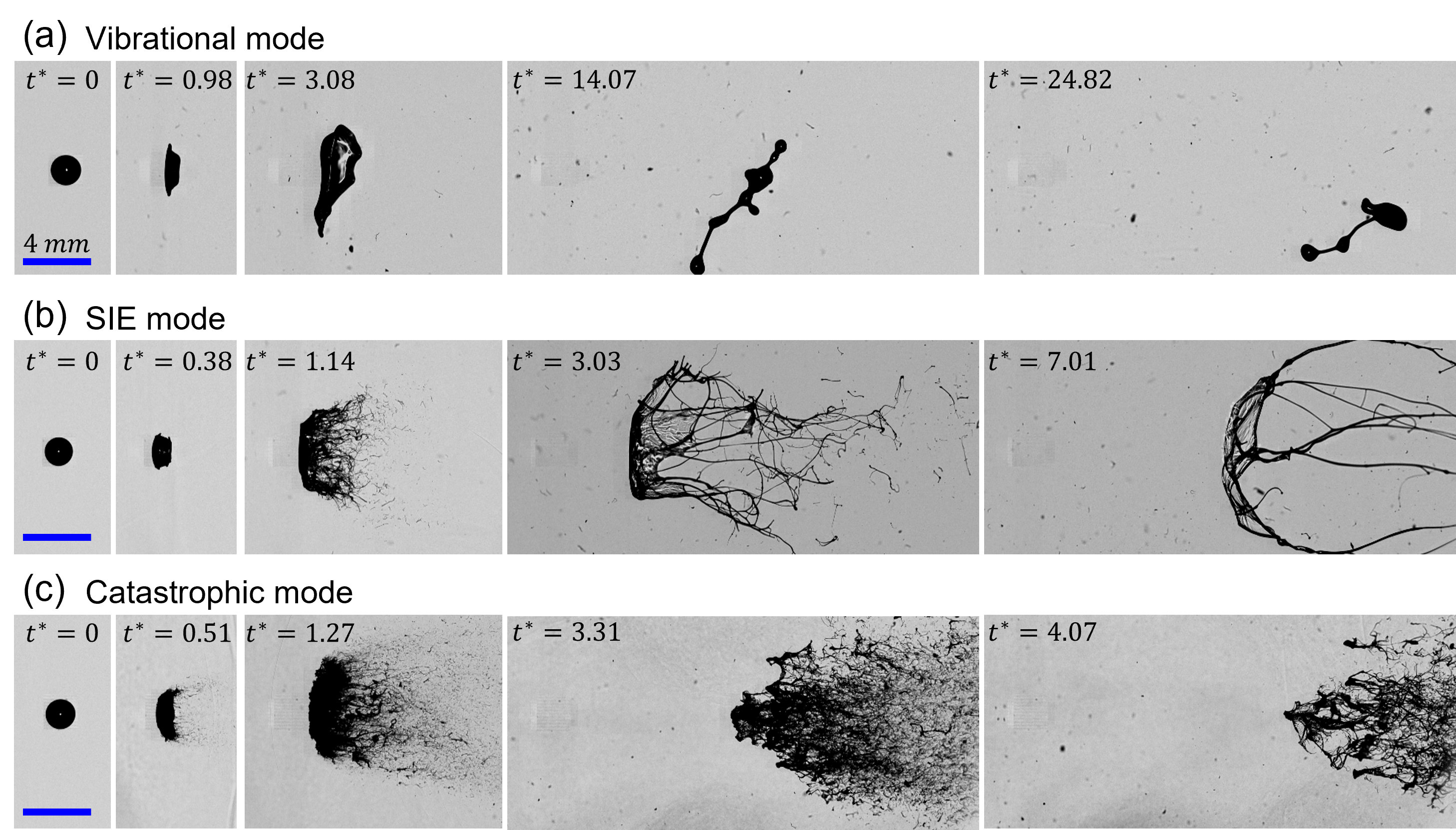}
\caption{Global observation on temporal evolution of polymeric droplet with $c/c^*=0.70$ subjected to $M_s$ of (a) 1.18, (b) 1.34 and (c) 1.71. Respective Weber numbers are 320, 949 and 7642. The three examples shown here represents the three modes of droplet breakup- Vibrational, SIE and catastrophic mode.}
\label{fig:global_obs}
\end{figure*}
Figure \ref{fig:global_obs} shows global observation on the temporal evolution of liquid droplet with $c/c^*=0.70$ subjected to $M_s$ of 1.18, 1.34 and 1.71 with respective $We$ as 320, 949 and 7642. Here, $t^*=t/t_I$ is the non-dimensionalized time such that $t_I=\frac{D_0}{U_g}\sqrt{\frac{\rho_l}{\rho_g}}$ is the inertial time scale, and time, $t$ is counted from the moment of droplet-shock wave interaction. The three cases shown in Figure \ref{fig:global_obs} illustrates the three different breakup modes observed in the present study (supplementary movie 1, 2 and 3). At low $We$, droplet suffers large deformation to form a flattened sheet ($t^*=3.08$ in Figure \ref{fig:global_obs}a). Finally, the surface tension and the elastic forces (if present) overwhelms the aerodynamic forces leading to rebound and oscillation of the liquid mass. This is identified as the vibrational mode of breakup (Figure \ref{fig:global_obs}a). In this regime, generally the liquid mass remains a single integral structure, and even if it breaks up only few daughter droplets are formed. Liquid rebound is the main characteristics of the vibrational mode. At moderate $We$, in addition to droplet deformation, KH waves also appear on the droplet surface  between the front stagnation point and the droplet equator. These waves grows with time and travels towards the equator region, finally leading to the ejection and liquid mass entrainment in the airflow near the equator region ($t^*=1.14$ in Figure \ref{fig:global_obs}b). This is identified as the SIE mode of breakup. In this regime the liquid mass drawn in the airflow goes beyond the rebound limit of restoring forces and this features separates it from the vibrational mode observed at low $We$.  At high $We$, in addition to droplet deformation and KH waves, the RT waves also contribute to the breakup process. RT waves appears as the surface corrugations on the flattened frontal region of the droplet ($t^*=1.27$ in Figure \ref{fig:global_obs}c), leading to a widespread erratic breakup of liquid mass which is identified as the catastrophic breakup mode (Figure \ref{fig:global_obs}c).

\subsection*{Stages of droplet aerobreakup}
\begin{figure*}
\centering
\includegraphics[width=1.0\linewidth]{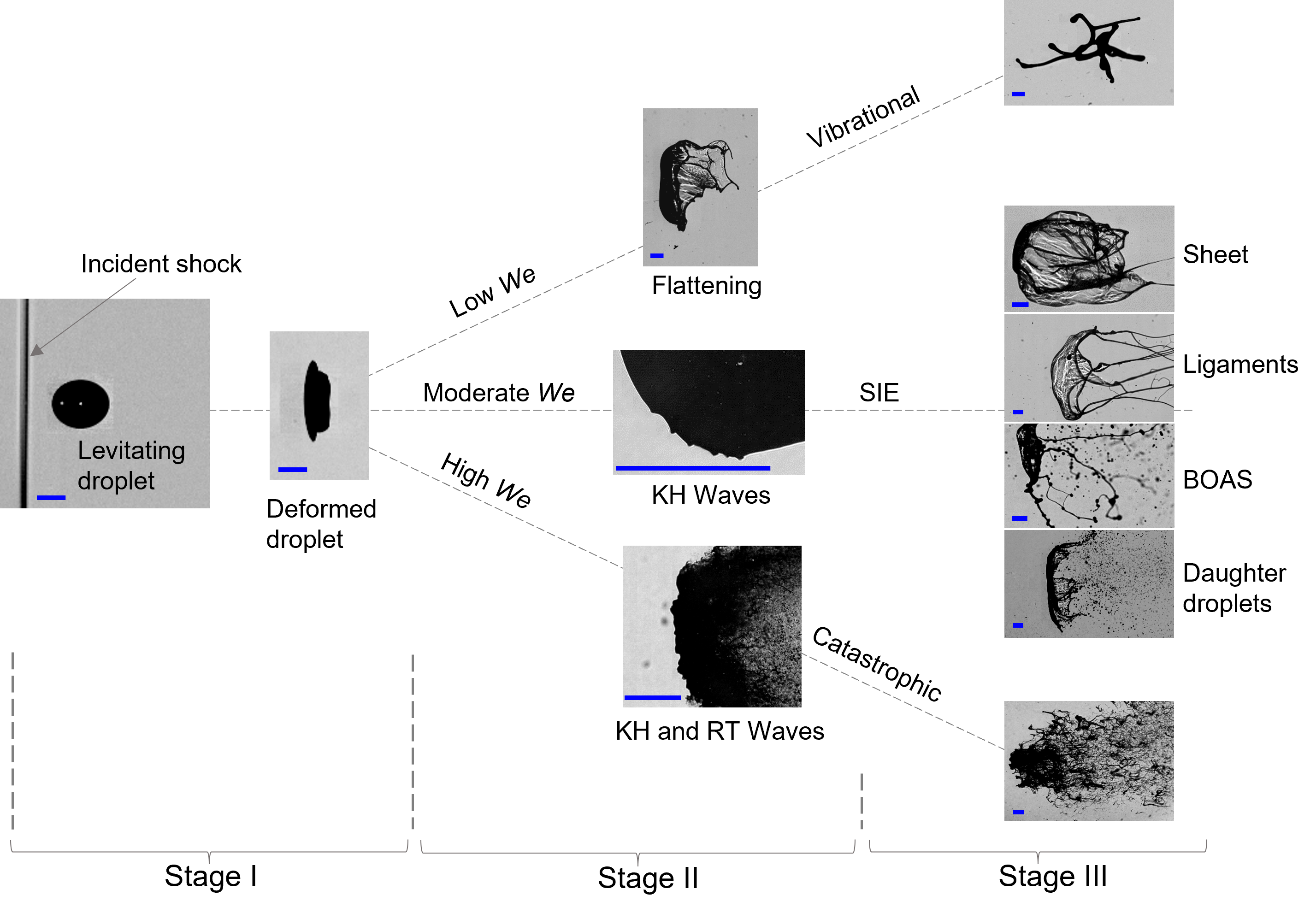}
\caption{The three stages in droplet aerobreakup based on the temporal evolution of the liquid mass. Scalebar in all the images represents 1 mm.}
\label{fig:stages}
\end{figure*}
On the basis of the temporal evolution of the liquid mass, we describe each mode of droplet aerobreakup as a three-stage process (supplementary movie 4). Figure \ref{fig:stages} illustrates the three stages with representative experimental images showing the state of liquid mass in each stage. Stage-I is the droplet deformation which occurs immediately after the passage of shock wave at the droplet location. In this stage cross stream diameter of the droplet increases and the shape of the droplet changes from spherical to a cupcake geometry. Stage-II marks the appearance and the growth of different hydrodynamic instabilities. State of the liquid mass obtained in the stage-II act as a precursor for the stage-III where breakup modes are decided and morphological evolution of liquid mass is observed. In stage-II and stage-III, three different regimes are observed based on the $We$. At low $We$ (<800), a deformed droplet undergoes further deformation to form a flattened sheet. Sometimes small ligaments are also observed emanating from the periphery of the flattened sheet. Finally the liquid mass is pulled back leading to the vibrational mode of breakup. At moderate $We$ ($800<800<2800$), droplet undergoes KH wave-assisted SIE mode. At high $We$ (>2800), RT wave-assisted catastrophic mode of breakup is observed. A clear-cut demarcation between the three stages is not always possible, and overlap of two consecutive stages also happens. At high $We$ regime, due to higher growth rate and quick appearance of hydrodynamic instabilities, it appears that the stage-I and the stage-II proceeds simultaneously. Capturing them separately is beyond the scope of the present temporal resolution. It is important to provide stage-wise description of the aerobreakup process because the effect of liquid elasticity is insignificant in the first two stages. Dominant role of elasticity appears only in the stage-III in terms of morphology of the liquid mass. In this stage, depending upon liquid elasticity, different liquid morphologies like, sheet, ligaments, bead-on-a-string (BOAS) etc. can be observed. Detailed discussion on each stage is provided in the following subsections.

\subsection*{Stage-I: Droplet deformation}
\begin{figure*}
\centering
\includegraphics[width=0.8\linewidth]{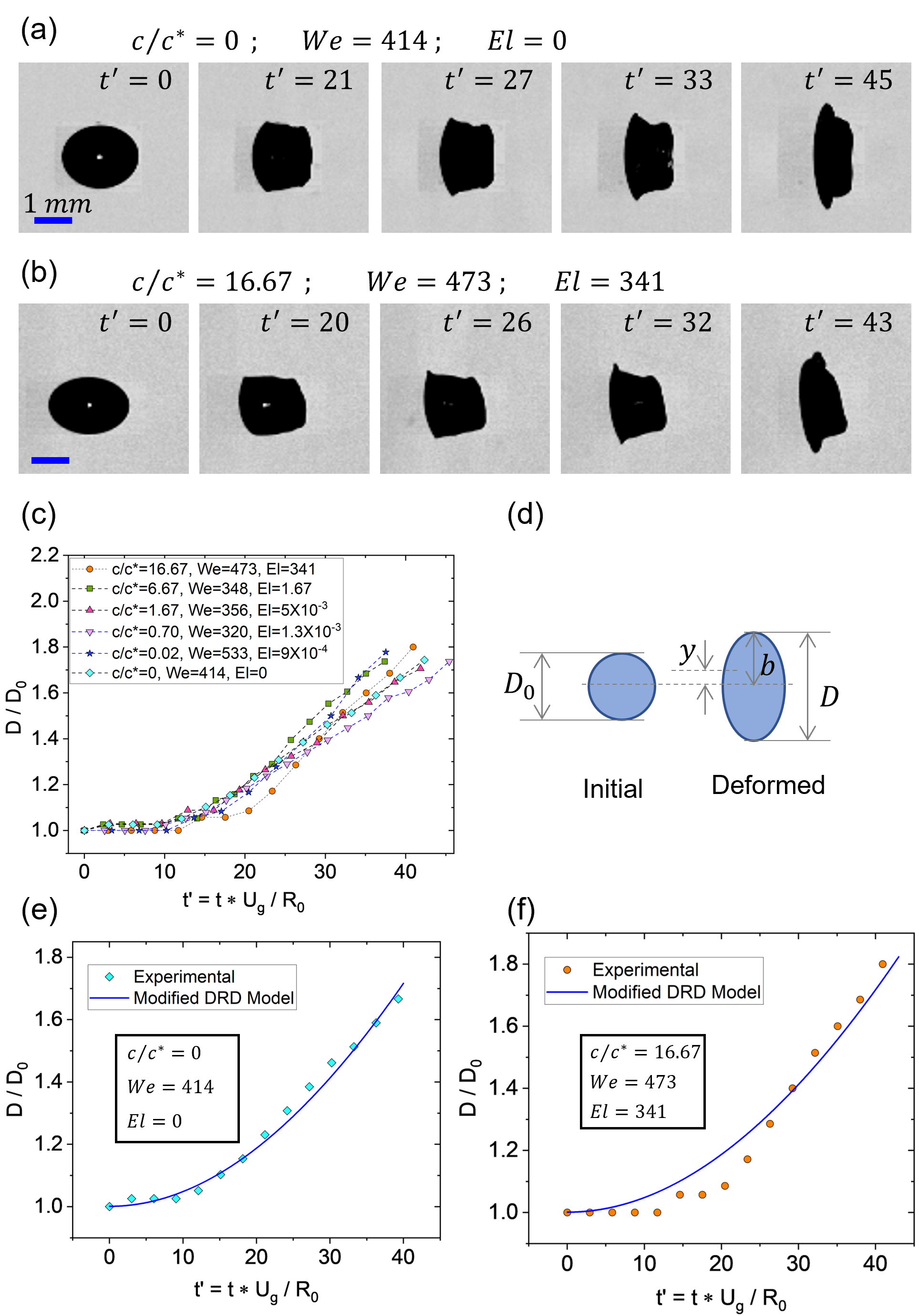}
\caption{Effect of liquid elasticity on the droplet deformation dynamics. (a,b) Experimental images showing aspect ratio evolution for water and polymeric droplet with $c/c^*=16.67$  (c) Temporal variation of droplet deformation for water and polymeric droplets with different $c/c^*$ at comparable $We$. (d) Schematic of a spherical droplet deforming into the shape of an oblate spheroid. (e,f) Comparison of experimental deformation with the modified DRD model for $c/c^*$ values of 0 and 16.67 respectively.}
\label{fig:deformation}
\end{figure*}

At early times of droplet-airflow interaction, the air-stream almost achieves stagnation pressure at the windward and the leeward side of the droplet. Under action of this pressure, droplet deforms to change its shape from sphere to a cupcake geometry, often approximated to an oblate spheroid for modelling purpose \citep{sor2015modeling,sharma2021dynamics}. Deformation is quantified in terms of aspect ratio $D/D_0$ where $D$ is the maximum cross stream diameter of the deformed droplet. Figure \ref{fig:deformation}a and \ref{fig:deformation}b shows the temporal evolution of the aspect ratio for water and polymeric droplet with $c/c^*=16.67$ at comparable $We$. Deformation data for all concentrations is collated in \ref{fig:deformation}c. Here, $t' =tU_g/R_o$ is the non-dimensional time and $R_0$ is the initial radius of the droplet. Time, $t$ is counted from the instant of droplet-shock wave interaction and till $D/D_0 \sim 1.7$. It is clear Figure \ref{fig:deformation}a, b, and c that, variation of polymer concentration and hence $El$ have no significant effect on the deformation dynamics of the polymeric droplets.

A deforming droplet is assumed to undergo purely extensional flow, changing its shape from spherical to an oblate spheroid with semi-major axis length, $b$ as shown in the schematic diagram Figure \ref{fig:deformation}b. $y$ is the distance between half-droplet center of mass and the center of the spheroid such that $b=\frac{8}{3}y$. To predict the deformation, force balance is performed in terms of motion of the half-droplet centre of mass. \citet{sor2015modeling} proposed droplet ratio deformation (DRD) model to predict the deformation of Newtonian droplets. Here we present modified DRD model for polymeric droplet by adding the viscoelastic force term. The final force balance is given by Equation \ref{eqn:eqn1}
\begin{equation}
m\cdot a_c=F_v+F_s+F_p+F_{ve}
\label{eqn:eqn1}
\end{equation}
Here, $m$ and $a_c$ are the mass and the acceleration of the half-droplet. Terms on the right hand side of Equation \ref{eqn:eqn1} represents the forces on half droplet due to viscous$(F_v)$, surface tension$(F_s)$, pressure$(F_p)$ and viscoelastic $(F_{ve})$ effects. It should be noted that $F_v$ is the viscous force only due to contribution from solvent (water in the present case), whereas $F_{ve}$ is the viscoelastic force due to polymer contribution. Except for $F_{ve}$, all the other terms in Equation \ref{eqn:eqn1} can be evaluated in the same manner as done in the literature \citep{sor2015modeling}. To estimate $F_{ve}$ we have used upper convected Maxwell (UCM) model for viscoelastic fluid subjected to a 2-dimensional incompressible and purely elongational flow. The expression obtained for $F_{ve}$ is given in Equation \ref{eqn:eqn2}
\begin{equation}
F_{ve}=\frac{2}{3}\pi R_0^3 \times \frac{\mu_p}{y\lambda}\xi
\label{eqn:eqn2}
\end{equation}
Where,
\begin{equation}
\xi=e^{-t/\lambda}\left[ \left( \frac{y}{y_0}\right)^2-\left( \frac{y_0}{y}\right)^2\right]  +\frac{e^{-t/\lambda}}{\lambda} \left[ y^2 \int_{0}^{t} \frac{e^{\tau/\lambda}}{y^2} \,d\tau  -  \frac{1}{y^2} \int_{0}^{t} y^2e^{\tau/\lambda}\,d\tau  \right]
\label{eqn:eqn3}
\end{equation}
$\mu_p$ is the polymer contribution to the zero shear viscosity of the polymeric solution, and $y_0=\frac{3}{8}R_0$ is the initial value of $y$. After substituting expression for all the terms in Equation \ref{eqn:eqn1} and performing suitable non-dimensionalization, the equation which governs the deformation is given by Equation \ref{eqn:eqn4}
\begin{equation}
\frac{d^2y'}{dt'^2} = -\frac{8N}{KRe} \left( \frac{1}{y'^2}\frac{dy'}{dt'}\right)   -  \frac{4}{KWe} \left( \frac{dA_s'}{db'}\right)  + \frac{3C_p}{4K} -  \frac{2M}{KReDe} \left( \frac{\xi}{y'}\right)
\label{eqn:eqn4}
\end{equation}
Where, $\xi$ in non-dimensionalized terms can be written as
\begin{equation}
\xi=e^{-t'/De}\left[ \left( \frac{y'}{y'_0}\right)^2-\left( \frac{y'_0}{y'}\right)^2\right]  +\frac{e^{-t'/De}}{De} \left[ y'^2 \int_{0}^{t'} \frac{e^{\tau/De}}{y'^2} \,d\tau  -  \frac{1}{y'^2} \int_{0}^{t'} y'^2e^{\tau/De}\,d\tau  \right]
\label{eqn:eqn5}
\end{equation}
Here, dashed quantities represent non-dimensional terms. $R_0$ and $R_0/U_g$ have been used as the length-scale and time-scale for non-dimensionalization. $C_p$ is the coefficient of pressure, $A_s$ is the surface area of half-spheroid and $A_s'=A_s/\pi R_0^2$. Non-dimensional numbers appearing in Equation \ref{eqn:eqn4} are density ratio ($K$), viscosity ratio($N$ and $M$), Reynolds number ($Re$) and Deborah number ($De$) defined as
\begin{equation}
\begin{split}
K=\frac{\rho_l}{\rho_g}; \hspace{0.7cm} N=\frac{\mu_s}{\mu_g}; \hspace{0.7cm} M=\frac{\mu_p}{\mu_g}; \hspace{0.7cm}
Re=\frac{\rho_g U_g D_0}{\mu_g}; \hspace{0.7cm} De=\frac{\lambda U_g}{R_0}
\label{eqn:eqn6}
\end{split}
\end{equation}
$\mu_g$ and $\mu_s$ are the dynamic viscosity of the gas phase and the solvent (water in the present case). Comparison of deformation predicted from Equation \ref{eqn:eqn4} with experimental data is shown in Figure \ref{fig:deformation}c and \ref{fig:deformation}d for two extreme values of $c/c^*$ studied in the present work. Here, $C_p$ is used as a fitting parameter, and from the different experimental data of deformation rate, its suitable value is obtained between 0.3 to 0.4. For the present range of parameters, it can be checked from Equation \ref{eqn:eqn4} that, contributions of viscous, viscoelastic and surface tension forces are at-least an order of magnitude less than the pressure term. This suggests that the droplet deformation is mainly governed by the balance between aerodynamic pressure and inertia of the deforming liquid. Since liquid properties play an insignificant role in the deformation dynamics, this explains that the temporal evolution of the aspect ratio is unaffected by the variation in $c/c^*$ of the test liquid (Figure \ref{fig:deformation}c). In non-dimensional form (Equation \ref{eqn:eqn4}), neglecting viscous, viscoelastic and surface tension terms leads to a constant deformational acceleration of the half-droplet. This means that the net deformation should be quadratic in time which is indeed observed experimentally as well as predicted theoretically (Figure \ref{fig:deformation}).

\subsection*{Stage-II: KH instability}
\begin{figure*}
\centering
\includegraphics[width=.8\linewidth]{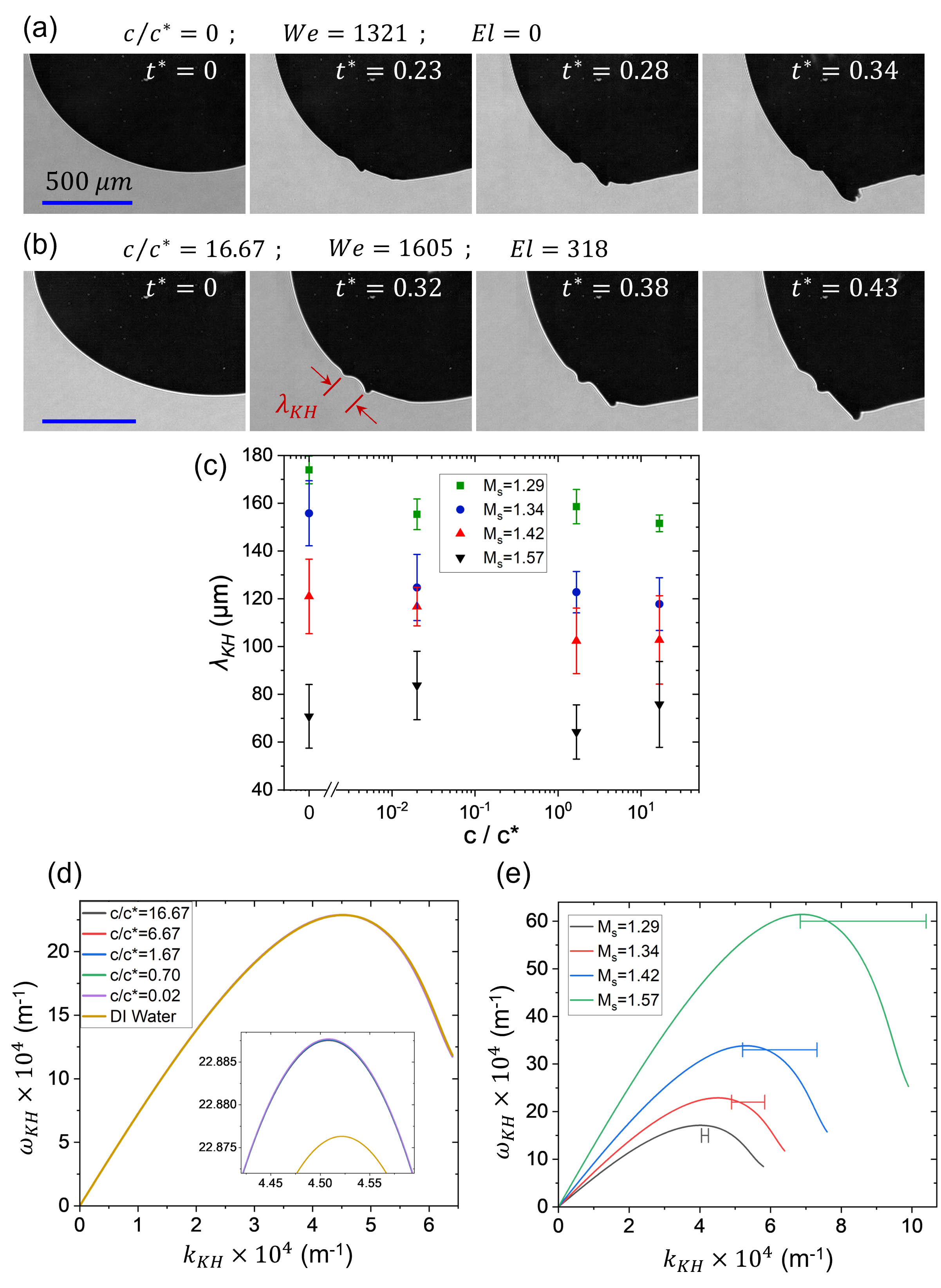}
\caption{Effect of liquid elasticity on the dynamics of KH instability. (a,b) Zoomed-in images showing KH waves on the surface of droplets with $c/c^*$ value 0 and 16.67 respectively. (c) Experimental value of KH wavelength for different concentrations at different shock mach number. (d) Dispersion plot of KH instability for different concentrations at $M_s=1.34$. Inset shows the dispersion plot near peak growth rate values. (e) Dispersion plot of KH instability for $c/c^*=16.67$ at different shock mach numbers. Horizontal line corresponds to the experimental wavenumbers with standard error as the span.}
\label{fig:KH_dynamics}
\end{figure*}

Stage-II represents the appearance and growth of the hydrodynamic instabilities. As shown in Figure \ref{fig:stages}, three different regimes based on $We$ can be observed in stage-II. With increasing $We$, the first transition ($We\sim800$) is observed when instead of only deformation, KH waves also appears on the droplet surface. To probe this further, KH instability wavelength, $\lambda_{KH}$ is measured for different cases. To measure $\lambda_{KH}$, high-speed imaging (75000 fps) at high resolution (4.5 $\mu$m / pixel) is performed as shown in Figure \ref{fig:KH_dynamics}a and \ref{fig:KH_dynamics}b. These experimental images shows the formation and evolution of KH waves for the case when a normal shock wave with $M_s=1.34$ is subjected to water and polymeric droplet with $c/c^*=16.67$ which is the maximum concentration considered in the present study. Only a quarter portion of the initially spherical droplet is captured to keep the high spatio-temporal resolution. Levitating droplet makes it convenient to keep the desired portion of the droplet in the camera field of view. The use of levitating droplets proves to be superior than the pendant droplet \citep{jackiw_ashgriz_2021} and the falling droplet method \citep{theofanous2013physics, arcoumanis1994breakup} used in the previous studies. Pendant droplets are not symmetric due to the presence of needle contact on one side, whereas in the case of falling droplet, it is difficult to synchronize everything to capture images at high speed and high resolution in a small field of view. Moreover, falling polymeric droplets may exhibit asymmetry due to the presence of a long liquid tail formed while detaching from the source needle. All these challenges are overcame by using a levitating droplet.

To understand the effect of liquid elasticity, $\lambda_{KH}$ is measured for water and polymeric droplets with three different concentrations subjected to four different $M_s$. The result is shown in Figure \ref{fig:KH_dynamics}c. Insignificant effect of $c/c^*$ variation in $\lambda_{KH}$ is observed for a given $M_s$. However, a monotonic decrease in $\lambda_{KH}$ with increasing $M_s$ is observed for all concentrations. Considering an inviscid and incompressible airflow on the droplet surface, neglecting the gravity effects and assuming a finite vorticity layer on gas phase, one can apply the linear perturbation analysis to obtain dispersion relation for KH instability as done for Newtonian droplets \citep{marmottant2004spray,padrino2006shear}. We extended this approach for viscoelastic liquids by using Oldroyd-B constitutive equation for the liquid phase given by
\begin{equation}
\textbf{T}+\lambda\stackrel{\kern0.3em\nabla}{\textbf{T}}  = 2\mu_0 \left( \textbf{D} + \lambda_r \stackrel{\kern0.3em\nabla}{\textbf{D}} \right)
\label{eqn:eqn7}
\end{equation}
Here, \textbf{T} and \textbf{D} are the stress and the strain rate tensor. A triangle on top denotes the upper convected derivative. $\lambda_r=\frac{\mu_s}{\mu_0}\lambda$ is the retardation time. A linear stability analysis for small perturbations in the liquid phase assuming no base flow is performed. Considering perturbation of the normal mode form with growth rate $i\omega_{KH}$, the relation between perturbed stress tensor, $\textbf{T}_\textbf{p}$ and strain rate tensor, $\textbf{D}_\textbf{p}$ can be written as Equation \ref{eqn:eqn8} \citep{awasthi2021kelvin}
\begin{equation}
\textbf{T}_\textbf{p} = 2\mu_0 \left( \frac{1+i\lambda_r\omega_{KH}}{1+i\lambda\omega_{KH}} \right) \textbf{D}_\textbf{p}
\label{eqn:eqn8}
\end{equation}
Equation \ref{eqn:eqn8} is similar to constitutive relation for a Newtonian fluid with an effective viscosity, $\mu_{eff}$, such that
\begin{equation}
\mu_{eff} = \mu_0 \left( \frac{1+i\lambda_r\omega_{KH}}{1+i\lambda\omega_{KH}}\right)
\label{eqn:eqn9}
\end{equation}
Replacing the Newtonian viscosity with $\mu_{eff}$ for liquid phase in the model proposed for Newtonian droplets \citep{padrino2006shear,marmottant2004spray}, the dispersion relation for KH instability in a viscoelastic fluid can be written as
\begin{equation}
e^{-2 \eta}=[1+(\Omega-\eta)] \left[\frac{\Phi+(\Omega+\eta)\left\{2 \hat{\rho}-(1+\hat{\rho})(\Omega+\eta)-(1+\hat{\mu}) \beta \eta^{2}\right\}}{\Phi+(\Omega+\eta)\left\{2 \hat{\rho}-(1-\hat{\rho})(\Omega+\eta)-(1-\hat{\mu}) \beta \eta^{2}\right\}}\right]
\label{eqn:eqn10}
\end{equation}
$\Omega=-2\omega_{KH}\delta/U_g$ is the dimensionless growth rate and $\delta$ is the boundary layer thickness in the gas phase. $\eta=k_{KH}\delta$ is the dimensionless wavenumber and $k_{KH}=2\pi/\lambda_{KH}$ is the wavenumber. $\hat{\mu}=\mu_g/\mu_{eff}$ and $\hat{\rho}=\rho_g/\rho_l$ are the viscosity and density ratios of gas and liquid phase. $\beta$ and $\Phi$ are defined as
\begin{equation}
\beta = \frac{i4\mu_{eff}}{\rho_l U_g \delta} ; \hspace{0.7cm}
\Phi = J\eta + \left( \frac{\hat{\rho}}{We_2} \right)\eta^3 + 2\eta^2\hat{\mu}\beta
\label{eqn:eqn11}
\end{equation}
Where,
\begin{equation}
We_2 = \frac{\rho_g U_g^2 \delta}{4\gamma}; \hspace{0.7cm}
J = \frac{4(\rho_l-\rho_g)g \delta}{\rho_l U_g^2}
\label{eqn:eqn12}
\end{equation}
Boundary layer thickness, $\delta$ in the gas phase is evaluated from the following expression
\begin{equation}
\frac{\delta}{D_o} \sim \frac{1}{\sqrt{Re_{D_0}}} ; \hspace{0.7cm} \frac{\delta}{D_o}= \frac{C}{\sqrt{Re_{D_0}}}
\label{eqn:eqn13}
\end{equation}
$Re_{D_0}$ is the Reynolds number in the gas phase with $D_0$ as the characteristic length scale. Proportionality constant, $C$ has been used as a fitting parameter. Equation \ref{eqn:eqn10} is solved numerically to plot the growth rate of KH instability corresponding to different wavenumbers. Figure \ref{fig:KH_dynamics}d shows the dispersion plot for water and polymeric droplets with different concentrations subjected to $M_s=1.34$. Dispersion curve for all the concentrations overlap for a given $M_s$. Near the peak point, the growth rate for water is slightly lower compared to the polymeric droplet as shown in the inset of Figure \ref{fig:KH_dynamics}d. This is because of the small difference in surface tension of water as compared to the polymeric solution. Figure \ref{fig:KH_dynamics}e shows the KH instability dispersion curve for a polymeric droplet with $c/c^*=16.67$ subjected to different $M_s$, and the horizontal line on the plot indicates the experimental value of wavenumbers with standard error as the span. It should be noted that the growth rate is not measured experimentally, and hence the horizontal lines are accurate only to the wavenumbers. It can be observed that the predicted wavenumber, $k_{KH_{max}}$ corresponding to the maximum growth rate of KH instability is in good agreement with the experimental values. It is clear from the experimental observations and theoretical predictions (Figure \ref{fig:KH_dynamics}) that the dynamics of KH instability is decided by the $M_s$ (and hence $We$), but the elasticity of the liquid phase plays an insignificant role.

\subsection*{Stage-II: RT instability}
\begin{figure*}
\centering
\includegraphics[width=.8\linewidth]{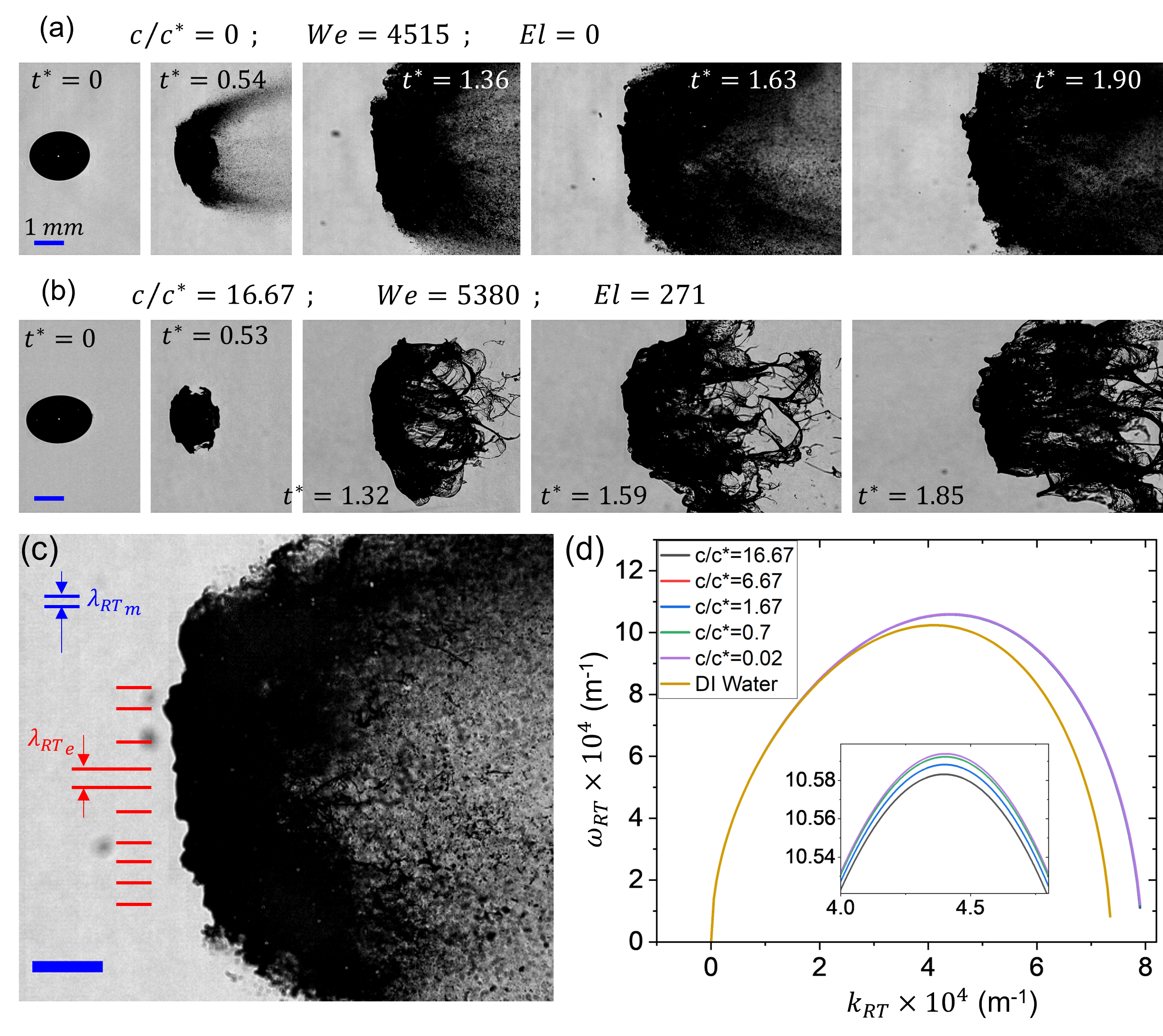}
\caption{Effect of liquid elasticity on the dynamics of RT instability. (a,b) Development of RT waves as surface corrugations on the flattened front surface of water droplet, and polymeric droplet with $c/c^*=16.67$ which is the maximum concentration considered in the present study. (c) RT waves on the front surface of polymeric droplet with $c/c^*=1.2\times10^{-3}$ at $M_s=1.57$, $We=5062$. Red lines indicates the wavelength of the surface corrugations i.e., the experimental value of RT wavelength, $\lambda_{RT_e}$. Two horizontal blue lines shows the RT wavelength, $\lambda_{RT_m}$ predicted by the theoretical model. (d) Dispersion plot of the RT instability for different concentrations at $M_s=1.57$. Inset shows the dispersion plot near peak growth rate values. Scalebar for all the experimental images corresponds to 1 mm.}
\label{fig:RTP}
\end{figure*}

As the $We$ is increased, second transition ($We\sim2800$) in stage-II is marked by the appearance of RT waves, in addition to the deformation and the KH waves (Figure \ref{fig:stages}). Again, the onset of critical $We$ for this transition is independent of $El$ similar to the first transition due to KH waves as discussed in the previous section. The RT waves appears as surface corrugations on the flattened front surface of the droplet as shown in Figure \ref{fig:RTP}a ($t^*=1.63$), \ref{fig:RTP}b ($t^*=1.59$) and \ref{fig:RTP}c. Figure \ref{fig:RTP}a and \ref{fig:RTP}b presents the development of RT waves on the surface of water droplet and high concentration polymeric droplet ($c/c^*=16.67$) both subjected to shock wave with $M_s=1.57$. For a given aerodynamic condition, the wavelength of front surface corrugations have similar magnitude irrespective of elastic properties. \citet{joseph2002rayleigh} provided theoretical analysis to relate these surface corrugations with the RT waves. Later, \citet{theofanous2008physics,theofanous2013physics} performed LIF imaging at an oblique angle of $30^{\circ}$ from the direction of side view imaging, and showed that the windward surface of the droplet remains smooth even at high $We$ condition. They claimed that the appearance of surface corrugation at high $We$ is an artifact due to the side view shadowgraphic imaging. Despite this ambiguity in existing literature, there is no denying from few facts. First, even with the shadowgraphic images, the transition from smooth to the corrugated front surface of the droplet is observed when the $We$ is increased from moderate to high value. Second, the RT waves do not appear as quickly as the KH waves due to large differences in their growth rate \citep{sharma2021shock}. Before the appearance of RT waves, the droplet has already suffered some deformation and growth of KH waves, resulting in the development of a significant cross-sectional area normal to the air flow direction. This liquid-air interface, perpendicular to the airflow direction provides a suitable platform for the RT instability to occur. Further, the cross-stream dimension of the flattened front surface is observed to be at least 1.3 times of $D_0$ at the time of appearance of corrugations, whereas \citet{theofanous2008physics} reported the span of front smooth surface to be only 0.65 times of $D_0$. This means that even if the RT waves do not form in the central portion of the front surface, there is enough normal surface area away from the central portion for the RT waves to form. This also suggests the possibility of hybrid KH-RT instability where RT waves are modulated on the crest of KH waves \citep{zhao2014influence,theofanousMitkin2017physics,sharma2021shock} and appears as the front surface corrugations.

In the previous section, we have already shown that the viscoelastic properties do not play a significant role in the KH instability for the considered range of parameters. Now we show that similar results are obtained for RT instability. We have used the dispersion relation for RT instability at high $We$ in an Oldroyd-B fluid derived by \citet{joseph2002rayleigh}. For all practical purpose $\rho_g<<\rho_l$ holds true, therefore neglecting $\rho_g/(\rho_l+\rho_g)$ and assuming $\rho_l/(\rho_l+\rho_g)=1$, the dispersion relation for RT instability with growth rate $i\omega_{RT}$ and wavenumber $k_{RT}$ can be written as
\begin{equation}
-1+ \frac{1}{\omega_{RT}^2}  \left( -ak_{RT}+\frac{\gamma k_{RT}^3}{\rho_l}\right) + \frac{4k_{RT}^2}{i\omega_{RT}} \left( \frac{\mu_g-\mu_{eff}}{\rho_l}\right)  - \frac{4k_{RT}^3}{\omega_{RT}^2} \left( \frac{\mu_g-\mu_{eff}}{\rho_l}\right)^2 \left( q_{2}-k_{RT}\right)=0
\label{eqn:eqn14}
\end{equation}
Where, 
\begin{equation}
q_2 = \sqrt{  k_{RT}^2 + \frac{i\omega_{RT}\rho_l}{\mu_{eff}} }
\label{eqn:eqn15}
\end{equation}
Here, $\mu_{eff}$ can be obtained from Equation \ref{eqn:eqn9} by replacing $\omega_{KH}$ with $\omega_{RT}$. Acceleration, $a$ of the droplet can be estimated from the following relation \citep{zhao2018transition}
\begin{equation}
a = \frac{3\rho_g U_g^2 C_d D_m^2}{4 D_0^3 \rho_l}
\label{eqn:eqn16}
\end{equation}
$C_d$ is the drag coefficient, and $D_m$ is the maximum cross-stream diameter of the liquid mass. The RT instability depends strongly on the acceleration of the air-liquid interface. For an exact estimation of acceleration, in Equation \ref{eqn:eqn16}, $U_g$ should be replaced by relative velocity between the air and the droplet, which changes continuously with time. $C_d$ depends on $D_m$ which in turn is a transient quantity. This makes it difficult to obtain the exact value of acceleration. However an approximate estimate can be obtained using Equation \ref{eqn:eqn16} and taking $C_d=1.2$ and $D_m=2.15D_0$ \citep{zhao2018transition}. Now, Equation \ref{eqn:eqn14} can be solved numerically to plot the dispersion curve for RT instability. In the original formulation, Joseph et al. have used retardation time as a fitting parameter to match the theory with experiments. However, retardation time is not an independent parameter for an Oldroyd-B fluid. It depends on the relaxation time, zero shear viscosity of the solution and the solvent viscosity such that $\lambda_r=(\mu_s/\mu_0)\lambda$ \citep{larson2013constitutive}. Using the correct value of $\lambda_r$, Equation \ref{eqn:eqn14} is solved to get the RT dispersion plot for all the test liquid droplets subjected to $M_s=1.57$ equivalent to $We \approx 4900$ for polymeric droplet with $D_0=1.8$ mm. The results are shown in Figure \ref{fig:RTP}d. Inset shows the dispersion plot near peak growth rate values. It can be observed that the dispersion curve overlaps for all the polymeric liquids. Water shows a slight deviation from the polymeric liquids, mainly because of the small difference in the surface tension values. Wavelength of RT instability, $\lambda_{RT_{m}}$ (blue) predicted from model is shown in Figure \ref{fig:RTP}c for comparison with the experimental data, $\lambda_{RT_{e}}$ (red). Due to various approximations in the theory, especially in estimating the acceleration, $a$, this model is good only for an order of magnitude estimation of RT wavelengths. Despite its drawbacks, this model brings out the essential physics that the RT instability is insensitive to the liquid elasticity for a wide range of values studied in the present work.

\subsection*{Stage-III: Breakup morphology}
\begin{figure*}
\centering
\includegraphics[width=1\linewidth]{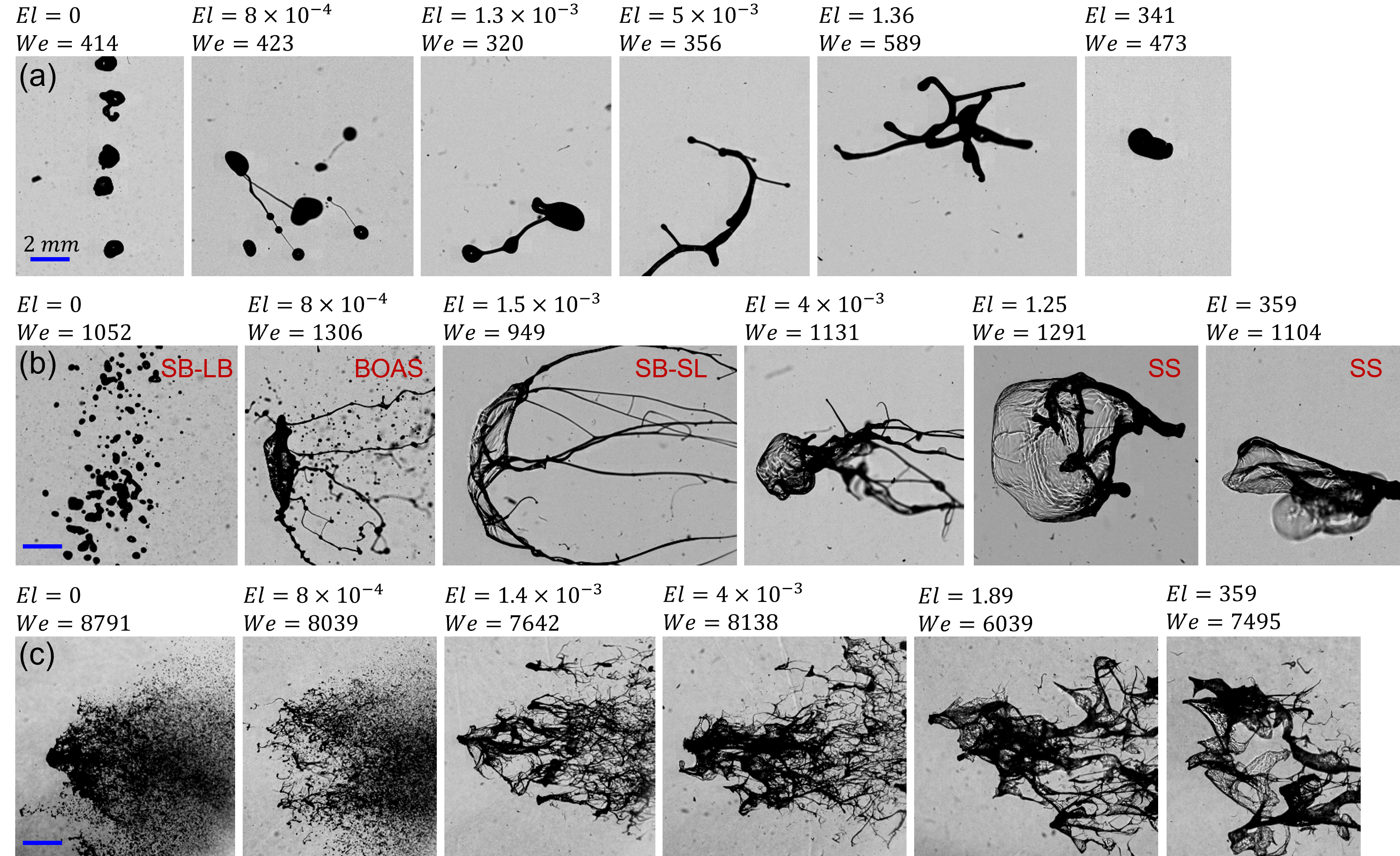}
\caption{Effect of $El$ on morphology of liquid mass obtained in stage-III of (a) Vibrational mode, (b) SIE mode and (C) catastrophic breakup mode. Well defined morphologies obtained in stage-III of SIE mode are denoted as- SB-LB: sheet breakup ligament breakup, BOAS: bead on a string, SB-SL: sheet breakup stable ligament, SS: stable sheet.}
\label{fig:morphology}
\end{figure*}

From the perspective of viscoelastic droplet aerobreakup, stage-III is the most important stage where the effect of the liquid elasticity enters into the play. This stage corresponds to the evolution of liquid mass morphology pertaining to different modes of aerobreakup. Three distinct modes are identified in the present study as shown in stage-III of the Figure \ref{fig:stages}. At low $We$, droplet undergoes large deformations and subsequent rebound of the liquid mass leading to the vibrational mode of breakup. At moderate $We$, droplet undergoes KH wave-assisted SIE mode, and at high $We$, the droplet experiences RT wave-assisted catastrophic mode. These breakup modes are decided by the underlying hydrodynamic instabilities during stage-II. We have already shown in the previous subsections that the liquid elasticity plays an insignificant role in the first two stages, therefore the critical $We$ for the onset of different breakup modes are also independent of $El$. However, the effect of $El$ is seen during stage-III in terms of the liquid mass morphology obtained within a particular mode of breakup (Figure \ref{fig:morphology}). 
Increase in $El$ caused higher degree of rebound in the vibrational regime, and larger size of fragmented chunks in the catastrophic regime. Effect of elasticity is even more prominent in the stage-III of the SIE mode. In this regime, based on $El$, a hierarchy of well defined morphological structures are observed. At high  $El$, the liquid mass entrained in the airflow forms a  stable sheet denoted as SS in the Figure \ref{fig:morphology}b. As the $El$ number is lowered, sheet structure is not stable, they break, and long stable ligament structures are observed denoted as SB-SL (sheet breakup-stable ligament). In the case of water, ideally $El=0$, neither sheet nor ligament structure are stable, and they break to form fine daughter droplets denoted as SB-LB (sheet breakup-ligament breakup). Another interesting observation is the formation of a bead on a string structure (BOAS), reported widely in extensional flows of dilute polymeric solutions \citep{clasen2006beads,bhat2010formation,malkin2014polymer,scharfman2016visualization}. BOAS consist of small beads of liquid mass connected by threadlike ligaments, as shown in the Figure \ref{fig:morphology}b. BOAS is observed for moderate $We$ number (SIE regime) and lower range of $El$ number. Below this $El$, ligaments are not stable and form fine droplets through the mechanism of Rayleigh-Plateau instability (RPI). Above this $El$, stable ligaments are formed. BOAS is the intermediate morphology, presenting the transition from the stable ligament to the fine daughter droplet formation. 

\begin{figure*}
\centering
\includegraphics[width=.8\linewidth]{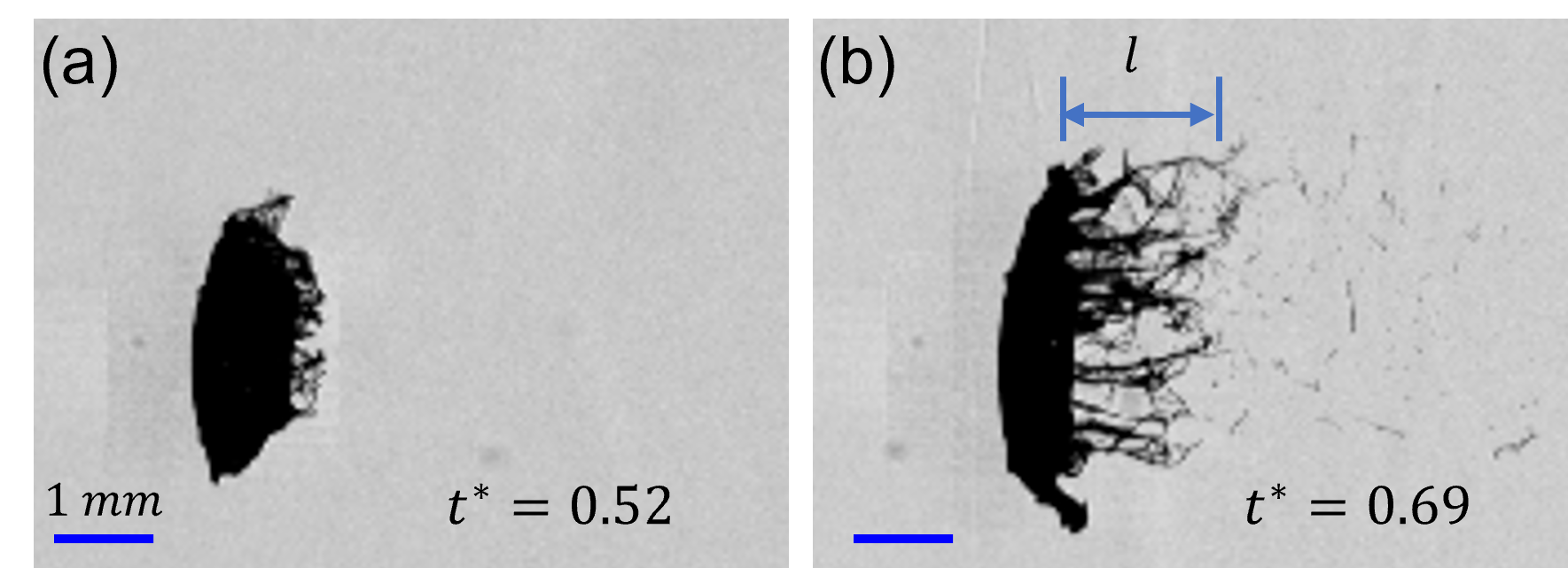}
\caption{Strong extensional flow near the equator region of a polymeric droplet with $c/c^*=1.67$, $El=4\times10^{-3}$ undergoing SIE mode of breakup at $We=987$.}
\label{fig:extensional}
\end{figure*}

All these morphologies observed in stage-III shows that the liquid elasticity enters as a stabilizing agent or resistance against the aerobreakup process. It is a challenging task to theoretically model these complex phenomena, especially the fragment size at high $We$. There are theoretical studies related to the breakup of viscoelastic sheets and ligaments, but the results are ambiguous in a sense that, some literature reported elasticity as a destabilizing agent \citep{liu1998linear,wang2015weakly,dasgupta2021effects}, whereas others reported it as a stabilizing agent \citep{goren1982surface,bousfield1986nonlinear,ruo2011three,yang2013instability,xie2016effects}. The resolution for this discrepancy comes in terms of unrelaxed tension \citep{xie2016effects} in the liquid phase. The role of elasticity changes from destabilizing to stabilizing agent if the unrelaxed tension is accounted in the relevant stability analysis. Unrelaxed tension develops in strong extensional flows of polymeric solution, where flow timescale is small compared to $\lambda$, such that it does not allow enough time for relaxation of polymer molecules \citep{bousfield1986nonlinear}. The condition for the presence of unrelaxed tension can be given in terms of the Weissenberg number, $Wi=\lambda\dot{\varepsilon}>1$, where $\dot{\varepsilon}$ is the strain rate of extensional flow. Various literature mentioned above, provide stability analysis for viscoelastic sheets and ligaments in the context of flow coming out from a spray nozzle. The same physics can be extended to the stability of morphologies obtained in the SIE regime of the present study. In this regime, the liquids mass entrained in the airflow suffers strong extensional flow as it is getting drawn into sheets and ligaments (Figure \ref{fig:extensional}). 
An order of magnitude estimation of $\dot{\varepsilon}=\frac{1}{l}\frac{dl}{dt}$ can be obtained from the images as shown in Figure \ref{fig:extensional}, where $l$ is the instantaneous length of liquid mass entrained in the airflow. Average values of $\dot{\varepsilon}$ pertaining to SIE regime is in the range of $\sim10^4 s^{-1}$. The value of $\lambda$ for the polymeric solutions used in the present study is in the range of $\sim10^{-3}-10^{-1} s$. Here, $Wi>>1$ suggest the presence of unrelaxed tension in the liquid phase and hence, stabilizing role of elasticity. This explains the increased stability of liquid morphologies with increasing $El$ as observed in the stage-III of the present study.

\subsection*{Regime map of droplet breakup}
Traditionally, aerobreakup modes of Newtonian droplets are presented in a $We-Oh$-number space \citep{guildenbecher2009secondary}. A similar plot, by replacing $Oh$ with an effective Ohnesorge number, $Oh_{eff}$, has been used for non-Newtonian droplets \citep{theofanous2013physics,theofanousMitkin2017physics}. Drawback of $Oh_{eff}$ is that it only accounts for the shear thinning/thickening effect through liquid viscosity but not the elastic effects. $We$ represents the ratio of aerodynamic force to the surface tension forces, and it is equally important for aerobreakup of viscous and viscoelastic droplets. Whereas, $Oh$ brings in the effect of liquid viscosity which is relevant only for viscous droplets. In case of the viscoelastic liquids, apart from viscosity and surface tension, elasticity also plays an important role. The elasticity of any liquid appears in terms of a non-zero finite relaxation time, $\lambda$. It is obvious that a  non-dimensional number like $El$ that accounts for both viscous and elastic effects must be chosen in the case of viscoelastic droplet breakup, which has been not done before. $El$ represents the relative strength of viscoelastic force to the inertial force in the liquid phase, and it can be estimated from the Equation \ref{eqn:eqn00}. $El$ has been shown relevant for understanding the primary breakup of viscoelastic liquid sheets and ligaments \citep{thompson2007atomization,wang2015weakly}. Such structures are also observed in the present study (Figure \ref{fig:morphology}b) in a way that the morphological hierarchy is in coherence with $El$. This motivates the usage of $El$ as a crucial governing parameter in the present work.
\begin{figure*}
\centering
\includegraphics[width=1.0\linewidth]{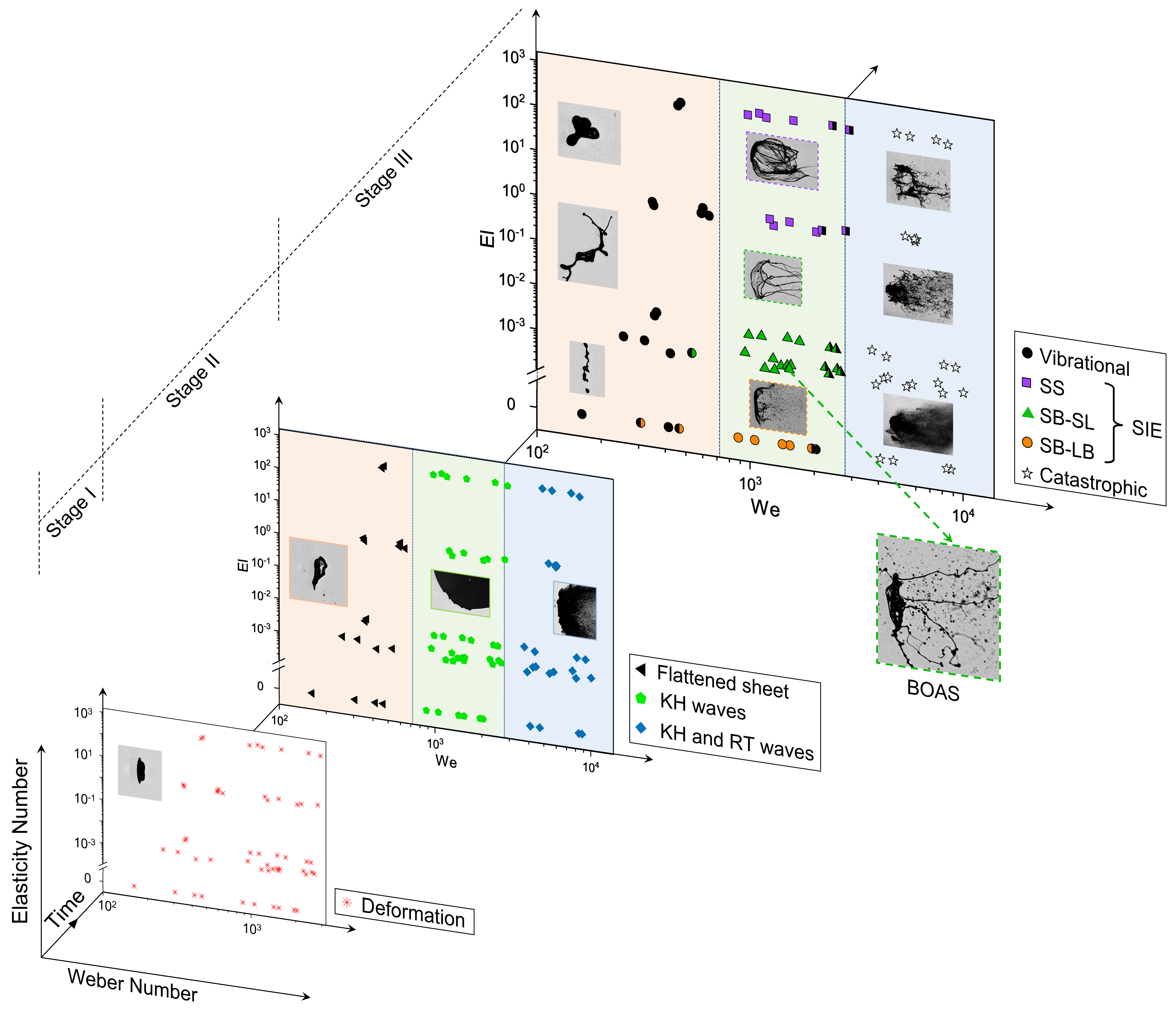}
\caption{3-dimensional regime map showing different stages and breakup modes of viscoelastic droplet breakup on a $We-El-time $ axes. Symbols with color partition indicates transition between two consecutive modes.}
\label{fig:regimemap}
\end{figure*}

Results obtained from all the experimental runs are plotted in the form of a 3D phase plot in $El-We-time$ coordinate as shown in Figure \ref{fig:regimemap}. Representative experimental images are shown as inset in the plot. Droplet deformation (stage-I) happens for complete range of $We$ and $El$ considered in the present study. At high $We$ (>2800) conditions, stage-I and stage-II seems to proceed simultaneously therefore, these data points are not shown in stage-I. The stage-II and the stage-III is divided into three distinct regimes. Orange region corresponds to low $We$ ($<800$) regime where unstable flattened sheet formed in stage-II leads to vibrational mode of breakup in stage-III. Green region indicates the moderate $We$ ($800<We<2800$) regime where KH instability in stage-II leads to SIE breakup mode in stage-III. Blue region corresponds to high $We$ ($>2800$) regime where RT instability in stage-II causes catastrophic breakup in stage-III. The critical $We$ for transition from one regime to next regime is independent of $El$ and this is supported by the experimental evidences and mathematical analysis provided in the previous subsections. Inset images provided in the stage-III shows the stabilizing effect of $El$ in the aerobreakup process. Well defined liquid morphologies obtained in SIE regime are shown with different symbols.

\section{Conclusions}
\label{sec:Conclusions}
Present exposition outlined the role of elasticity on the underlying mechanism for different stages in aerodynamic breakup of viscoelastic droplets. Multi-order variation in governing non-dimensional parameters- $We$ $(\sim10^2-10^4)$ and $El$ $(\sim10^{-4}-10^2)$ is investigated in a precise experimental arrangement. This provides an excellent benchmark, much needed for future studies in the less explored area of viscoelastic droplet breakup. Three distinct breakup modes are identified with increasing $We$ on a $We-El$ plane, which are- vibrational ($We<800$), KH wave-assisted SIE ($800<We<2800$) and RT wave-assisted catastrophic mode ($We>2800$). Each mode can be described as a three-stage process based on the temporal evolution of the liquid mass. Stage-I is the droplet deformation, stage-II corresponds to the development of different hydrodynamic instabilities, and stage-III involves the morphological evolution of liquid mass undergoing a particular mode of breakup. At high $We$, stage-I and the stage-II proceeds simultaneously due to higher growth rate of hydrodynamic instabilities. It is interesting to note from the present experiment and the supporting mathematical analysis that, the liquid elasticity plays an insignificant role in the stage-I and II of the droplet aerobreakup. Since, breakup modes are decided by the governing hydrodynamic instability in stage-II therefore, the boundaries of $We$ for the onset of different breakup modes are independent of $El$. Significant effect of elasticity appears only during the stage-III of the droplet breakup in terms of higher stability of liquid structures (or retardation to the breakup) with increasing $El$. Especially in SIE regime, a hierarchy of well defined liquid structures (sheet, ligament, BOAS, daughter droplets) are obtained with variation in $El$. Strong extensional flows in stage-III ensures the presence of unrelaxed tension in the liquid phase, which in turn causes the elasticity to act as a stabilizing agent against the breakup of liquid structures. This provides direction for future theoretical works that the unrelaxed tension must be accounted in the breakup of viscoelastic droplets.

\section*{Acknowledgments}
The authors acknowledge support from IGSTC (Indo–German Science and Technology Center) through project no. SP/IGSTC-18-0003. N.K.C. acknowledges support from the Prime Ministers Research Fellowship (PMRF).

\section*{Supplementary movies}\label{sec:sup_vid}
The supplementary movie files as supporting material are also provided.

\textbf{Movie 1}: vibrational breakup mode.

\textbf{Movie 2}: SIE breakup mode.

\textbf{Movie 3}: Catastrophic breakup mode.

\textbf{Movie 4}: The three stages of droplet breakup.

\section*{Declaration of Interests}

The authors report no conflict of interest.

\section*{Author ORCIDs.}
Navin Kumar Chandra https://orcid.org/0000-0002-1625-748X;\\
Shubham Sharma https://orcid.org/0000-0002-8704-887X;\\
Saptarshi Basu https://orcid.org/0000-0002-9652-9966;\\
Aloke Kumar https://orcid.org/0000-0002-7797-8336.

\nocite{*}
\bibliography{aipsamp}

\end{document}